\documentclass[
  reprint,
  superscriptaddress,
  amsmath,amssymb,
  aps,
  prapplied,
  floatfix
]{revtex4-2}

\usepackage{graphicx}
\usepackage{dcolumn}
\usepackage{bm}
\usepackage{graphicx}
\usepackage{xcolor}
\usepackage{braket}
\usepackage[icon=Note,width=0.12em,height=0.12em,opacity=0.5,color=yellow]{pdfcomment}
\usepackage{orcidlink}
\usepackage[normalem]{ulem}
\newcommand{\ra}[1]{\textcolor{black}{#1}}
\newcommand{\rb}[1]{\textcolor{black}{#1}}
\newcommand{\rc}[1]{\textcolor{black}{#1}}
\newcommand{\rd}[1]{\textcolor{black}{#1}}
\begin{document}

\title{Photon-echo synchronization and quantum state transfer in short quantum links}%

\author{Hong Jiang\orcidlink{0009-0008-7937-6671}}
\email{These authors contributed equally to this work.}
\affiliation{Instituto de Física Fundamental, IFF-CSIC, Calle Serrano 113b, Madrid 28006}
\affiliation{Quantum Advanced Research Center (QuARC-CSIC), Calle Serrano 113b, Madrid 28006}

\author{Carlos Barahona-Pascual\orcidlink{0009-0007-2372-9397}}
\email{These authors contributed equally to this work.}
\affiliation{Instituto de Física Fundamental, IFF-CSIC, Calle Serrano 113b, Madrid 28006}
\affiliation{Quantum Advanced Research Center (QuARC-CSIC), Calle Serrano 113b, Madrid 28006}

\author{Juan José García-Ripoll\orcidlink{0000-0001-8993-4624}}
\email{jj.garcia.ripoll@csic.es}
\affiliation{Instituto de Física Fundamental, IFF-CSIC, Calle Serrano 113b, Madrid 28006}
\affiliation{Quantum Advanced Research Center (QuARC-CSIC), Calle Serrano 113b, Madrid 28006}

\date{\today}
\begin{abstract}
  The short quantum link regime, where the photon travel time $\tau$ is comparable to the emitter lifetime $1/\gamma$, is experimentally relevant but theoretically underexplored: existing few-mode descriptions lose validity as retardation and multimode effects become significant. Using a Delay Differential Equation (DDE) framework that admits exact analytical solutions from the single-mode cavity limit to the multimode waveguide continuum, we show that emitters coupled to a short link spontaneously lock into self-synchronized Rabi oscillations driven by coherent photon echoes, breaking the link's discrete time-displacement symmetry. The resulting spectral structure---persistent quasi-dark states and vacuum Rabi splitting, including in the superstrong coupling regime---enables efficient quantum state transfer (QST): benchmarking three protocols across the full $\gamma\tau$ parameter space, we find that STIRAP exploits the quasi-dark-state structure to achieve a quadratic infidelity floor $\mathcal{O}((\gamma\tau)^2)$, outperforming both SWAP (linear error $\mathcal{O}(\gamma\tau)$) and wavepacket engineering for $\gamma\tau \lesssim 1.44$, even in regimes where retardation cannot be neglected. These results establish photon-echo synchronization as an engineering resource for quantum state transfer, with DDE modeling providing the exact analytical predictions needed to design and optimize short-link experiments on current circuit-QED hardware.
\end{abstract}

\maketitle

\section{Introduction}

High-fidelity transfer of quantum states between distant nodes is a central requirement of quantum networks~\cite{kimble2008,awschalom2021,zheng2026}. This operation can be realized deterministically using quantum optical networks based on superconducting transmission lines~\cite{pechal2014,zhong2019,magnard2020,chang2020,conner2021,penas2022,he2025}, photonic waveguides~\cite{sipahigil2016,zheng2026}, and optical fibers~\cite{serafini2006,chen2007,zhou2009,vogell2017}.

Two extreme regimes bracket the physics of quantum optical interconnects, \rb{as quantified by the dimensionless product $\gamma\tau$, where $\gamma$ is the effective emission rate into the line and $\tau$ is the photon propagation time between emitters}. In cavity-like links, $\gamma\tau\ll 1$, the qubit interaction is mediated by a single discrete field mode and is effectively instantaneous, and the system is well described by a single-mode Tavis--Cummings model, enabling QST protocols such as SWAP~\cite{serafini2006} and stimulated Raman adiabatic passage (STIRAP)~\cite{chen2007,zhou2009,chang2020}. In waveguide-like links, $\gamma\tau\gg1$, the field forms a quasi-continuum and the finite photon travel time $\tau$ between emitters becomes a defining dynamical feature. This limit admits an input--output formulation~\cite{zhang2013,christiansen2026}, which forms the basis of QST via photon-shaping protocols such as the Cirac--Zoller--Kimble--Mabuchi (CZKM) scheme~\cite{cirac1997}.

However, the crossover through the intermediate regime $\gamma\tau\approx 1$, where a finite photon travel time coexists with a still-discrete mode structure, remains far less explored.
Recent circuit-QED experiments already operate in this regime: the 0.73~m link of Chang \textit{et al.}~\cite{chang2020} ($\gamma\tau\approx0.2$), the 5~m cable of Magnard \textit{et al.}~\cite{magnard2020} ($\gamma\tau\approx 1$), and the 64~m cryogenic link of Qiu \textit{et al.}~\cite{qiu2025} ($\gamma\tau\approx2$) all lie in this parameter space.
A prevailing intuition is that the dynamics is governed by only a few dominant waveguide modes, motivating single- or few-mode descriptions~\cite{vogell2017, malekakhlagh2024, he2025,zhang2025}.
Such treatments have two important limitations. First, the effective number of modes depends sensitively on the coupling strength, severely restricting their universality. Second, these truncations often overlook retardation effects and the intrinsically discontinuous dynamical features that emerge~\cite{parker1987,dorner2002,alvarez-giron2024}. %

This work explores the physics of short quantum links $\gamma\tau\lesssim1$ using the framework of Delay Differential Equations (DDE). We show that, \rb{in the limit in which the photon carrier frequency $\Delta$ is far from the bottom of the spectrum, $\Delta\gg1/\tau$}, this method is valid across a full range of regimes, from the single-mode cavity limit to the waveguide quasi-continuum. DDEs naturally incorporate retardation and encapsulate the entire spectral content in a compact set of functional equations, admitting exact analytical solutions for both the orbital dynamics and the spectral structure---eigenenergies and vacuum Rabi splitting---all the way to the so-called \textit{superstrong coupling} regime~\cite{meiser2006}, $\gamma\tau \gtrsim 1$, recently observed~\cite{kuzmin2019,johnson2019}.

In the short-link regime, where the single dimensionless parameter $\gamma\tau$ governs the crossover from cavity-like to retardation-dominated dynamics, we find that a single emitter undergoes photon-echo synchronization: it locks into piecewise non-differentiable Rabi oscillations driven by coherent photon echoes, a discrete time-crystal-like spontaneous breaking of the link's $2\tau$ time-translation symmetry without any external drive.

The same photon-echo synchronization mechanism exists for two emitters in a quantum link, where it becomes a resource that enables the implementation of efficient, high-fidelity quantum state transfer. Specifically, we demonstrate that protocols originally developed for cavities \cite{serafini2006,chen2007,zhou2009,vogell2017,ban2018,chang2020,maleki2021,malekakhlagh2024}, can be adapted to the short-link regime, where they leverage the persistent quasi-stationary states with large vacuum Rabi splittings across the full cavity-to-waveguide crossover.

We have developed and exhaustively benchmarked a complete toolbox of QST protocols across the full $\gamma\tau$ parameter space, deriving exact analytical error bounds and regions of optimality. A cavity-inspired SWAP protocol directly converts the joint non-differentiable Rabi oscillation into state transfer, with an infidelity that scales as $\mathcal{O}(\gamma\tau)$. A STIRAP-like protocol exploits the adiabatic quasi-dark-state structure to achieve a quadratic error floor $\mathcal{O}(2\times10^{-5}(\gamma\tau)^2)$, making it the protocol of choice throughout the short-link regime. Both STIRAP and SWAP are more efficient than the Cirac-Zoller-Kimble-Mabuchi (CZKM) protocol~\cite{cirac1997} up to the crossover $\gamma\tau\approx1.44$, where CZKM becomes the fastest strategy. The region of parameters where STIRAP is preferred becomes even larger if we consider photon losses in the link.

The remainder of the paper is organized as follows. Section~\ref{sec:model} derives the DDE framework for one and two emitters. Section~\ref{sec:single_qubit} analyzes the single-qubit synchronization dynamics and spectral structure. Section~\ref{sec:qst} extends these results to two-qubit quantum state transfer, benchmarking three protocols and establishing the superiority of STIRAP in the short-link regime. We conclude in Sec.~V with a summary and mapping onto current circuit-QED hardware.

\section{Delay Differential Equation framework\label{sec:model}}

\begin{figure}[t]
  \centering
  \includegraphics[width=\linewidth]{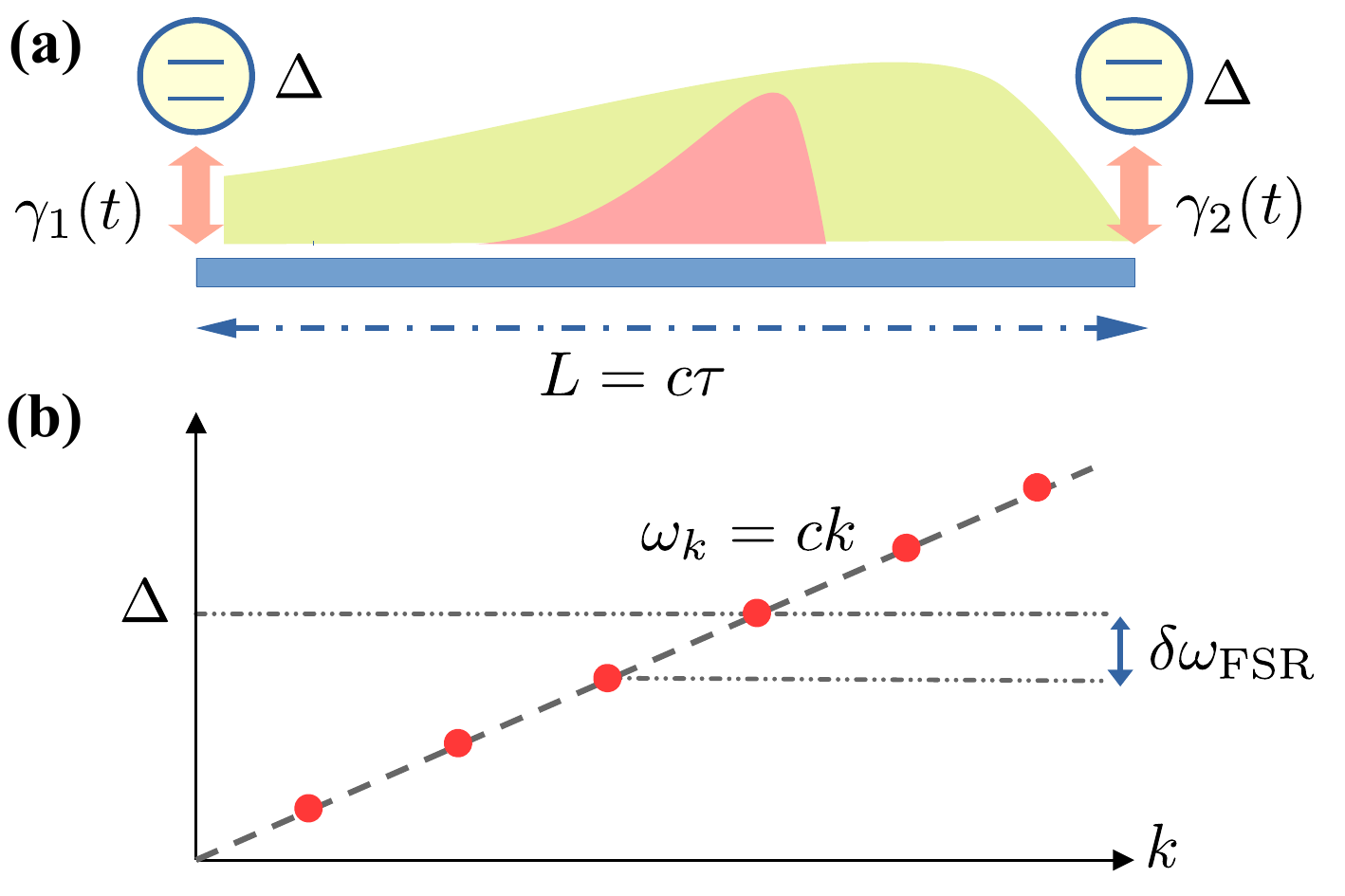}
  \caption{Minimal quantum link.
    (a) Two qubits interact with a closed waveguide of length $L$, with possibly time-dependent \rb{effective emission rates} $\gamma_{1,2}(t)$.
  (b) The waveguide supports an approximately linear dispersion $\omega_k=ck$, with a free spectral range $\delta\omega_\text{FSR}=\pi c/L=\pi/\tau$ determined by the time $\tau$ a photon takes to travel from one of the waveguide to another.\label{fig:setup}}
\end{figure}

Our goal in this section is to derive a set of Delay Differential Equations (DDEs) that accurately describe the dynamics of one or two quantum emitters coupled to a finite-length waveguide. Specifically, we consider the setup sketched in Fig.~\ref{fig:setup}. These are two identical two-level emitters with transition energy $\hbar \Delta$ placed at opposite ends of a cable of length $L$.
The field in the cable has a linear dispersion relation $\omega_{k} = c|k|$ and a discrete set of frequencies with a free spectral range $\delta\omega_\text{FSR} = \pi c/ L := \pi / \tau$, where we define $\tau = L/c$ as the time photons require to cover the cable's length.

The interaction between the emitters and the field is given by a dipole coupling under the Rotating Wave Approximation.
The Hamiltonian of this system $(\hbar = 1)$ is given by:
\begin{align}
  H =&  \sum_{l=1}^{2}\Delta \sigma^{+}_{l}\sigma^{-}_{l} + \sum_{k}\omega_{k}a_{k}^{\dagger}a_{k} +  \sum_{k,l} \left[  g_{k,l}(t)a_{k}\sigma^{+}_{l} + \mathrm{H.c.}\right], \label{eq:Hamiltonian}
\end{align}
where $g_{k,l}(t)$ is the time-dependent coupling between the $l$-th emitter and the $k$-th mode.

The Wigner-Weisskopf (WW)ansatz is an exact parameterization of all states in the single excitation sector of the previous Hamiltonian, and is given by
\begin{equation}
  \ket{\psi(t)} = \left[\sum_{l=1}^{2}c_{l}(t)\sigma^{+}_{l} + \sum_{k}\alpha_{k}(t)a_{k}^{\dagger}\right] \bigotimes_{l=1}^{2} \ket{g}_{l} \otimes \ket{\text{vacuum}}, \label{eq:ww_ansatz}
\end{equation}
where $|c_{l}(t)|^{2}$ is the probability for $l$-th qubit to be excited, and $\alpha_k$ describes the photons in the link.

It is possible to simulate exactly the dynamics of this quantum state for a relatively large number of modes. This will be indeed our baseline for all comparisons in the following sections. However, it is physically more interesting to derive an effective description of the dynamics in terms of the emitter coefficients $c_l(t)$ alone, which is the goal of the DDE framework.

\ra{In the WW simulations, we vary the mode truncation between $N_{\mathrm{mode}}=100$ and $400$, depending on the parameter regime. The finite number of modes affects mainly the Lamb shift and, to a lesser extent, the effective decay rate. For a specified spectral cutoff, both quantities converge as the number of retained modes is increased, and all WW results reported below were checked for convergence with respect to this truncation.}

Provided $g_{k,l}(t)$ varies slowly, we can follow the derivation in~\cite{barahona-pascual2025} to obtain scalar DDEs for the coefficients $(c_{l}(t))$.
These equations provide an accurate description when the bandwidth of the interaction between the emitters and the field is narrow enough that the integration limits in frequency can be taken to $ \pm \infty$.
This approximation is valid when the link is large compared to the photon wavelength, that is $\Delta / \delta\omega_\text{FSR}$ is large. However, as we will see below, these numbers do not have to be very large for the bandwidths we are considering $\gamma \lesssim 1/\tau$.

The resulting DDEs, in the frame rotating with the emitters' frequency read
\begin{align}
  \frac{d c_{l}(t)}{dt}  =& -\frac{\gamma_{l}(t)}{2}\,c_{l}(t) \label{eq:dde_two_emitter}  \\
  &-\sqrt{\gamma_{l}(t)} \big[e^{i2\phi}b_{l}^\text{out}(t-2\tau)  +
  e^{i\phi}  b_{3-l}^{\text{out}}(t-\tau) \big], \notag
\end{align}
The first term on the right-hand side of~\eqref{eq:dde_two_emitter} describes a local decay into the link with time-dependent rate $\gamma_{l}(t)/2$. The second term is the non-Markovian echo $b^\text{out}$ of all photons generated by both emitters in the past
\begin{equation}
  b_{l}^\text{out}(t) =\sum_{n=0}^{\infty}e^{i2n\phi}\left.\sqrt{\gamma_{l}(t_n)} c_{l}(t_n)\Theta(t_n)\right|_{t_n=t-2n\tau}.
\end{equation}
The arrival of the photon echoes is guarded by Heaviside step functions $\Theta$, while $\phi = \Delta\tau$ is the phase accumulated by each photon during along a single traversal of the link. As discussed in Appendix~\ref{app_analytical_dynamics}, the solutions to Eq.~\eqref{eq:dde_two_emitter} admit analytical solutions for one and two emitters.

The single-emitter limit of Eq.~\eqref{eq:dde_two_emitter} (retaining only the self-echo term) already contains the complete physics of photon-echo synchronization analyzed in Sec.~\ref{sec:single_qubit}; the cross-emitter echo then provides the inter-qubit coupling that drives quantum state transfer in Sec.~\ref{sec:qst}.

\section{\label{sec:single_qubit} Photon-echo synchronization in a single quantum link}

\begin{figure}[t]
  \includegraphics[width=0.95\columnwidth]{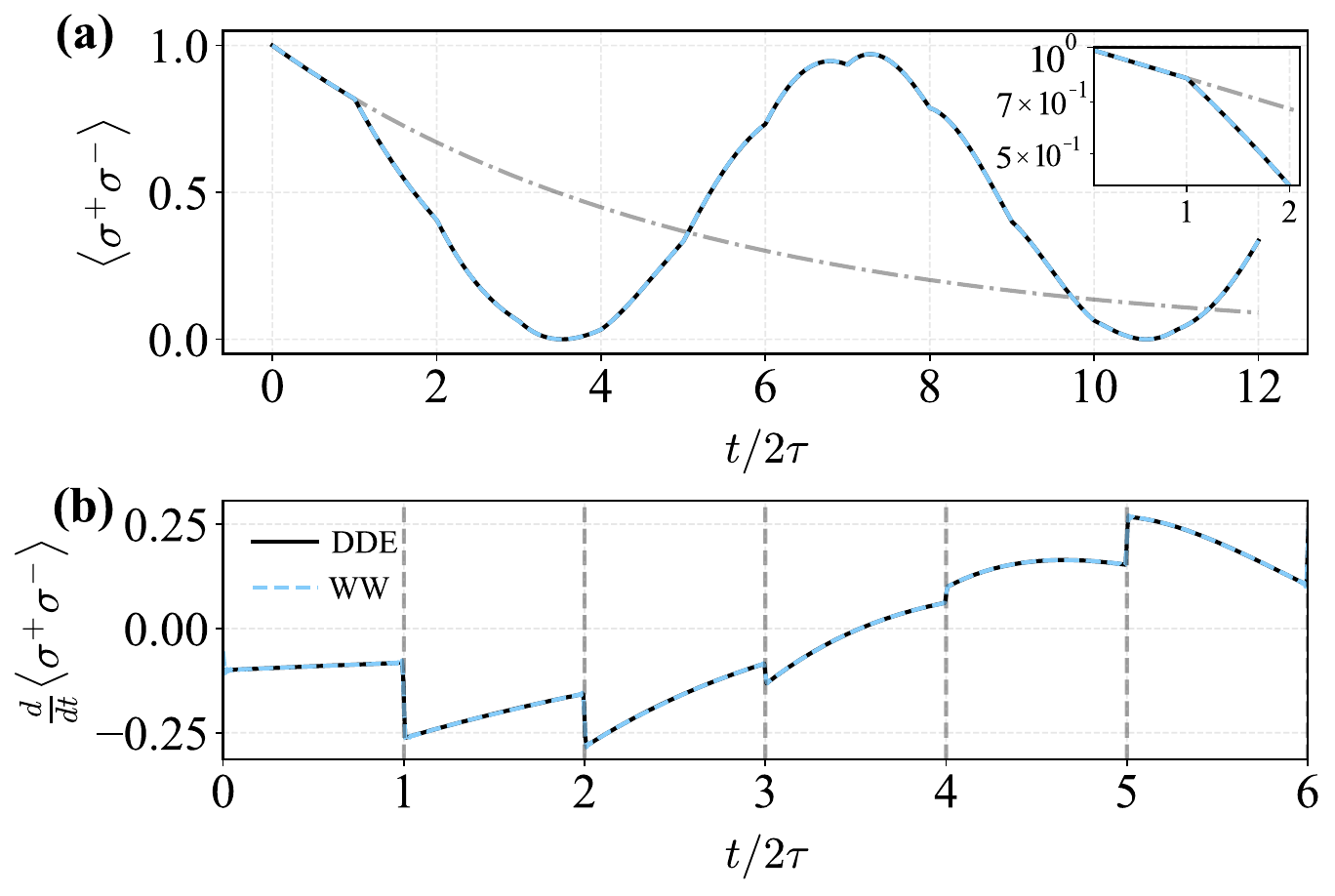}
  \caption{Dynamics of an initially excited two-level emitter in a medium-size cable.
    (a) Population of the excited state of the emitter as a function of time.
    (b) Time derivative of the population of the excited state for the first six periods of time.
    The solid black line represents the analytical solution of the DDE, and the dashed blue line shows the result of a numerical simulation using a Wigner-Weisskopf ansatz.
    The parameters taken were $\gamma \tau = 0.1$ and $\Delta = 50 \delta\omega_\text{FSR}$. \ra{$N_{\mathrm{modes}}= 100 $ for the WW simulation.}%
  \label{fig:osc_derivative}}
\end{figure}

In absence of external drives, the dynamics of a single excited emitter in the link is completely governed by the photon-echo mechanism.
We characterize this mechanism in two complementary ways.
Section~\ref{sec:dtc} focuses on the dynamics, revealing a photon-echo induced synchronization of the qubit, which develops a quasi-periodic piecewise non-differentiable Rabi oscillation with frequency $\Omega_R\approx\sqrt{\gamma/2\tau}$. We understand these oscillations as the emergence of periodic orbits $2\pi/\Omega_R$ that breaks the link's $2\tau$ discrete time-translation symmetry. This synchronized Rabi-like dynamics is the basis for the SWAP protocol from Sec.~\ref{sec:swap}.
Section~\ref{sec:non_diff_spectrum} focuses on the spectroscopy of the photons created during this quasiperiodic dynamics. An exact calculation reveals the emergence of well separated stationary states and vacuum Rabi splitting across the full coupling range, including the \textit{superstrong coupling regime}~\cite{meiser2006}. This spectral structure with large eigenenergy spacings will be the one exploited by the quasi-adiabatic STIRAP protocols from Sec.~\ref{sec:stirap}


\subsection{Symmetry-breaking Self-synchronization\label{sec:dtc}}
\begin{figure}[t]
  \includegraphics[width=0.95\columnwidth]{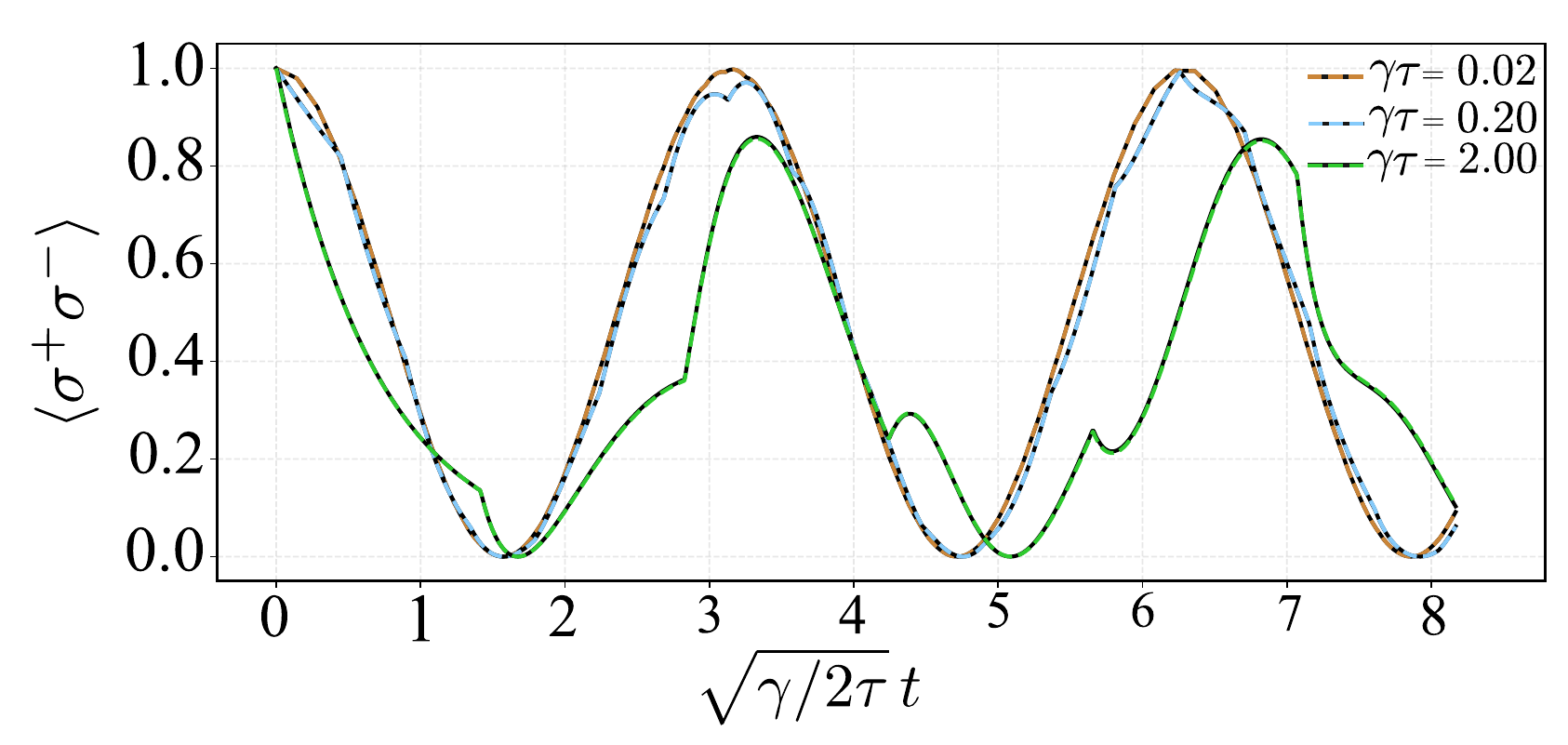}
  \caption{Dynamics of an initially excited two-level emitter in a medium-size cable for different values of $\gamma \tau$. The solid black lines correspond to the solution of the DDE, while the dashed lines represent the solution using a truncated Hamiltonian. \ra{$N_{\mathrm{modes}}= 150 $ for each truncated-Hamiltonian curve.}}%
  \label{fig:cavity_waveguide_crossover}
\end{figure}

Let us assume a single emitter with frequency $\Delta$ and a constant \rb{effective emission rate} $\gamma$ into the waveguide. Tuning the emitter's frequency on resonance with one of the link's eigenmodes $\Delta\approx\omega_k$ facilitates the release of the emitter's energy as one propagating photon. Initially, the qubit's excitation probability decays exponentially with a rate $\gamma$ (See  Fig.~\ref{fig:osc_derivative}a), creating a similarly shaped wavepacket. At $t=2\tau$, the photon returns to the emitter, re-interacting with the qubit and inducing a discontinuity in the population's derivative (See Fig.~\ref{fig:osc_derivative}b). This and the subsequent photon echoes lock the emitter into a \textit{self-synchronized} dynamics, with sustained, piecewise-differentiable quasi-periodic oscillations, at a frequency $\Omega_R\simeq \sqrt{\gamma/(2\tau)}$ that is generally incommensurate with the $2\tau$ time-translation symmetry of the free link.

This physics realizes the far-from-equilibrium operational core of the time-crystal concept introduced by Wilczek~\cite{wilczek2012}: the spontaneous breaking of a discrete time-translation symmetry by quantum dynamics, without any external periodic drive. Our system circumvents no-go theorems that prevent time-translation symmetry breaking on the ground state by operating in a far-from-equilibrium subspace. In this region, the discrete time crystal (DTC) order~\cite{else2016,khemani2016}---persistent, autonomous, sub-harmonic oscillation of a physical observable---arises without Floquet drivings, directly from the interaction between one quantum of excitation and the infinite-dimensional photonic field of the link, and is witnessed by the qubit's probability $c(t)$ and the field it generates $b^\text{out}(t)$---the order parameters of the broken symmetry.

This self-synchronization mechanism is intrinsically multi-mode, as evidenced by the agreement between large simulations of the full dynamics (dashed blue line in Fig.~\ref{fig:osc_derivative}a) and the analytical solutions of the DDE (solid black line). Fortunately, the DDE shows that the complete dynamics is ruled by a single parameter $\gamma\tau$ that controls the cross-over from a Jaynes-Cummings model in the limit $\gamma\tau \to 0$ to the waveguide mode $\gamma\tau\gg 1$  ( Fig.~\ref{fig:cavity_waveguide_crossover}). It must be remarked, however, that in all these regimes the dynamic exhibits characteristic corrugations, with non-differentiable features that are upper-bounded by $2\gamma$, as discussed in Appendix~\ref{app_discontinuity_bound}.

This result, which is consistent with related theoretical observations in cavity-QED~\cite{milonni1974,parker1987,giesen1996a,guimond2016,das2025} and studies of open waveguide-QED setups~\cite{alvarez-giron2024,dorner2002}, is actually a feature that can be observed in actual experiments with superconducting qubits and microwave waveguides~\cite{chang2020, magnard2020, qiu2025}. In particular, those setups that can tune the \rb{effective emission rate} $\gamma$, such as Ref.~\cite{zhong2019}, may explore the full crossover from apparently smooth Rabi dynamics, to weak corrugations $\gamma\tau=0.02$ (brown  dashed line, Fig. ~\ref{fig:cavity_waveguide_crossover}) and to fully resolved kinks $\gamma\tau=2$ (green dashed line,  \rc{Fig.~\ref{fig:cavity_waveguide_crossover}}). This can be done both by observing the qubit's dynamics---e.g., abruptly disconnecting the qubit-waveguide coupling prior to the measurement---as well by tracking the field that leaks from the waveguide and measuring the microwave field~\cite{magnard2020}.

\subsection{Spectroscopy\label{sec:non_diff_spectrum}}
\begin{figure}[t]
  \centering
  \includegraphics[width=\columnwidth]{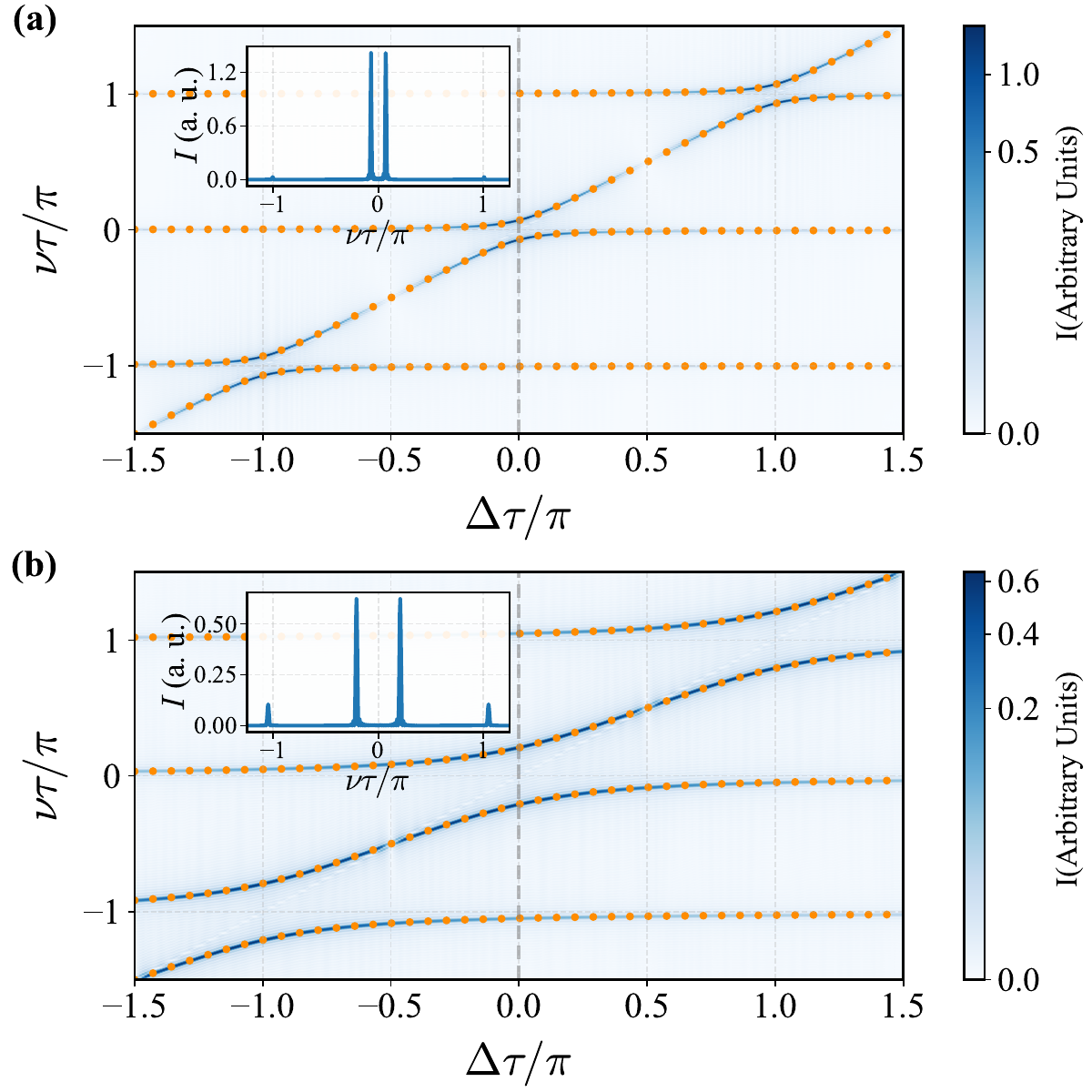}
  \caption{\ra{Power spectrum $I(\nu,\Delta)$ of the photons in the link. The horizontal axis is the normalized emitter detuning $\Delta\tau/\pi$, the vertical axis is the normalized spectral frequency $\nu\tau/\pi$, and the blue color scale gives the spectral intensity $I$ in arbitrary units.} In both cases, the orange dotted line is the analytical result of the eigenvalues, while the heat map is the result of a numerical analysis using the analytical solution.
    (a) weak coupling regime with $\gamma \tau = 0.15$.
  (b) superstrong coupling regime with $\gamma \tau = 1.5$. \ra{The insets show line cuts at the resonant setting $\Delta=0$, indicated by the vertical gray dashed lines in the main panels. Their horizontal axis is $\nu\tau/\pi$ and their vertical axis is the spectral intensity $I$ in arbitrary units.}  }%
  \label{fig:power_spectrum}
\end{figure}

The spectral analysis of the photons emitted by the qubit during this self-synchronized dynamics reveals a rich structure of eigenfrequencies and spectral weights that depend on the \rb{dimensionless parameter} $\gamma\tau$, and on the detuning $\Delta$ of the qubit from the quantum link's modes. These eigenfrequencies $\lambda$ can be computed analytically from the DDE (Appendix~\ref{app_eigenstates}), giving rise to the eigenvalue equation
\begin{equation}
  \lambda =\Delta  + \frac{\gamma}{2}\cot\left(\frac{\lambda\tau}{2}\right).
\end{equation}
This equation reveals a structure of well separated polaritonic branches and avoided level crossings, with a translational symmetry $\lambda\to\lambda+\delta\omega_\text{FSR}, \Delta\to\Delta+\delta\omega_\text{FSR}$, that is consistent with the approximation $\Delta/\delta\omega_\text{FSR}\gg 1$.

In the limit of moderate coupling $\gamma\tau \ll 1$ the analytical results confirm the predictions of a Jaynes-Cummings model, \rc{as illustrated in Fig~\ref{fig:power_spectrum}a)}. When the qubit approaches the resonance frequency of one waveguide mode ($\Delta\tau = \pi\times\mathbb{Z}$), we observe a vacuum Rabi splitting with the expected dependency $\delta \omega \approx 2 \sqrt{\gamma/2\tau}$ \rc{represented in the inset in Fig~\ref{fig:power_spectrum}a) for the case with $\Delta \tau = 0 $}. In contrast, the spectrum exhibits a pronounced suppression of spectral weight---a quasi-dark state---at the anti-resonance points $\Delta \tau = (2n-1)\pi/2$, where the qubit frequency is equidistant from two waveguide modes and the qubit's ability to decay or hybridize with the field is minimized. Both features are hallmarks of the strong coherent exchange between the emitter and the effective cavity formed by the cable, even though the complete dynamics still exhibits nonzero weights at other frequencies.

As the \rb{dimensionless parameter} $\gamma\tau$ increases (Fig.~\ref{fig:power_spectrum}b), the system enters a deeply non-Markovian regime, also referred as superstrong coupling~\cite{meiser2006, kuzmin2019, johnson2019}. In this limit, the \rb{effective emission rate} $\gamma$ becomes comparable to the free spectral range $\delta\omega_\text{FSR}=\pi/\tau$, and the qubit hybridizes with multiple modes of the link, leading to a more complex spectral structure. Despite this complication, the DDE still provides an exact analytical solution, revealing an enhancement of the vacuum Rabi splitting, whose growth necessarily falls below the single-mode prediction $2\sqrt{\gamma/2\tau}$, saturating to a fixed value $\pi/\tau$, in the limit $\gamma\tau\to+\infty$. In this limit participation of multiple modes becomes more evident in the Fourier transform, as shown in Fig.~\ref{fig:power_spectrum}b and the inset therein.

Despite the complexity of the spectrum, the combined photon-link system exhibits a robust ladder of hybridized states with large energy spacings $\mathcal{O}(\delta\omega_\text{FSR})$. These spacings suggest the possibility of inducing quasi-adiabatic protocols that exchange excitations between the emitters and the link they couple to. These quasi-adiabatic methods can operate at timescales $\mathcal{O}(1/\tau)$ and underly the most successful quantum state transfer methods in the following section.

\section{Quantum state transfer%
\label{sec:qst}}

The single-qubit self-synchronization mechanisms studied in Sec.~\ref{sec:single_qubit} extend naturally to two emitters connected by a short link.
The cross-emitter photon echo term $b_{3-l}^{\rm out}(t-\tau)$ in Eq.~\eqref{eq:dde_two_emitter} transmits the emission of one qubit to the other after a single traversal time $\tau$, producing a joint Rabi oscillation between the two emitters at frequency $\Omega_R\approx\sqrt{\gamma_0/2\tau}$---the same frequency identified in Sec.~\ref{sec:dtc} for the single-qubit self-echo---which can be used to exchange information between the quantum link's nodes.
Similarly, the semi-infinite ladder of hybridized qubit-photon stationary states discovered in Sec.~\ref{sec:non_diff_spectrum}, persists across the full cavity-to-waveguide crossover, suggesting STIRAP-like quasi-adiabatic protocols to exchange information while keeping the link nearly depopulated.

\rb{In this section, we exploit these two features---the joint photon-echo-induced Rabi oscillation and
the quasi-dark-state structure of the hybridized emitter--photon system---via two protocols}: one relies on always-on couplings that produce spontaneous synchronization (SWAP, Sec.~\ref{sec:swap}), and another that adiabatically engages the synchronization through smooth STIRAP-like controls (Sec.~\ref{sec:stirap}).

Both techniques operate well beyond the single-mode approximations used in the cavity-QED QST literature~\cite{serafini2006,chen2007,zhou2009,vogell2017,ban2018,chang2020,maleki2021,malekakhlagh2024} and outperform traditional quantum state transfer protocols based on wavepacket engineering.
We include the CZKM photon-shaping protocol~\cite{cirac1997} (Sec.~\ref{sec:czkm}) as a performance baseline to quantify this advantage and to identify the crossover beyond which wavepacket engineering becomes competitive.
Throughout, we work in the single-excitation sector; by linearity of the Schrödinger equation, the fidelity computed for the $|1\rangle$ component applies directly to arbitrary superposition states $\alpha|0\rangle+\beta|1\rangle$.

\subsection{SWAP: linear error scaling from coherent Rabi exchange%
\label{sec:swap}}

The ideal SWAP is inspired by the equivalent cavity-like regime protocol at $\gamma\tau \ll 1$, where the single-mode Tavis-Cummings model predicts a cavity-mediate coherent Rabi exchange of excitations between the two participating emitters\cite{serafini2006,conner2021}. Consequently, the equivalent protocol for the short quantum link is defined by two constant \rb{emission-rate profiles}
\begin{equation}
  \gamma^{\rm SWAP}_{1,2}(t)=\gamma_0, \quad t\in[0,T].
\end{equation}
\rb{where $T$ denotes the total duration of the state-transfer protocol.} Similar to Sec.~\ref{sec:dtc}, the two qubits couple to the link with an effective Jaynes-Cummings-like interaction $g\approx\sqrt{\gamma_0/2\tau}$. This suggests that Rabi oscillations happening at a frequency $\Omega=\sqrt{2}\Omega_R=\sqrt{2}g$ can implement a perfect swap of excitations at some resonant time $\Omega T\approx\pi$.

While this is strictly true in the limit $\gamma_0\tau\to0$, as we move away from the zero-length waveguide towards the short-link regime, both the non-differentiable multimode effects (Sec.~\ref{sec:dtc}) and the finite photon travel time distort the ideal cavity resonance condition. These effects cause the optimal QST duration to oscillate around the Rabi period and also reduce the overall transfer fidelity. \ra{Both predictions are confirmed by the DDE simulations presented in Appendix~\ref{app_swap}.}

Empirically, the minimal infidelity of the SWAP protocol follows a linear scaling,
\begin{equation}
  \varepsilon \sim \frac{3}{2}\gamma_0\tau,
\end{equation}
\ra{with an optimal duration $T_{\mathrm{SWAP}}/\tau\sim\pi/\sqrt{\gamma_0\tau}$; see
Appendix~\ref{app_swap} for details. SWAP thus performs optimally only in the quasi-cavity limit $\gamma_0\tau\ll1$, with the fidelity degrading rapidly as retardation effects grow.}

\begin{figure}[t]
  \centering
  \includegraphics[width=\columnwidth]{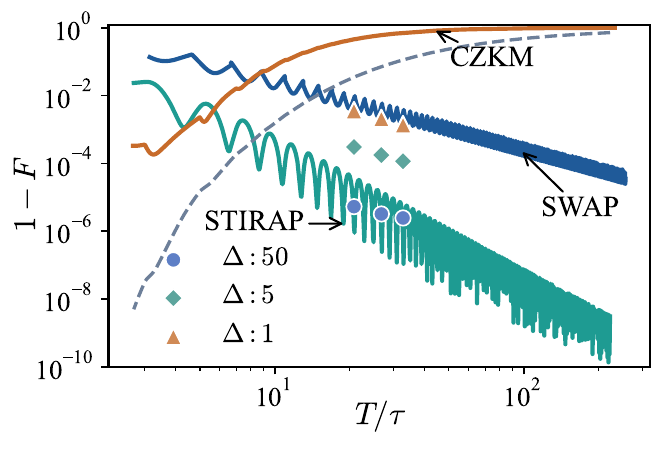}
  \caption{
    QST dynamics and efficiency comparison across the cavity-to-retardation crossover.
    \ra{
    Minimum infidelity $1-F$ for the three QST protocols as a function of their respective optimal dimensionless transfer durations $T/\tau$. Each protocol is evaluated at its own optimal $T/\tau$ (SWAP: $T/\tau\sim\pi/\sqrt{\gamma_0\tau}$; STIRAP: $T/\tau\sim9/\sqrt{\gamma_0\tau}$;
    CZKM: $T/\tau=9/\sqrt{\gamma_0\tau}$), so the comparison is performed at different absolute durations.}
    The dashed line denotes the exact CZKM lower bound on the infidelity.
    \ra{The nine discrete symbols are selected renormalized WW benchmark results for the STIRAP protocol, corrected for the Lamb shift at finite detuning: three representative points are shown for each of $\Delta/\delta\omega_{\mathrm{FSR}}=50$ (circles), $5$ (diamonds), and $1$ (triangles). $N_{\mathrm{modes}}=201$ for the WW benchmark points.} }%
  \label{fig:figure_qst}
\end{figure}

\subsection{STIRAP: quadratic error suppression via quasi-dark states%
\label{sec:stirap}}

While SWAP harvests the joint Rabi oscillation directly, STIRAP takes a more controlled approach: a pulse sequence adiabatically engages the inter-emitter coupling, steering the state through the eigenstate manifold and keeping the link’s electromagnetic field closer to vacuum than for other protocols\cite{chen2007,zhou2009,vogell2017,ban2018,chang2020} (c.f. Sec.~\ref{sec:losses}).
The quasi-dark states of the single-qubit spectral analysis (Sec.~\ref{sec:non_diff_spectrum}) persist into the two-qubit problem, enabling this adiabatic passage across both the cavity-like and short-link regimes and decoupling the protocol from the multimode distortions that limit the SWAP. \ra{In the cavity-like limit, the instantaneous quasi-dark state reduces to the familiar zero-photon expression
\begin{align}
\ket{D(t)} \approx
    \frac{
    \sqrt{\gamma_2(t)}\,\ket{eg;\text{vacuum}}
    -
    \sqrt{\gamma_1(t)}\,\ket{ge;\text{vacuum}}
    }{
    \sqrt{\gamma_1(t)+\gamma_2(t)}
    } .
\end{align}
It continuously evolves from $\ket{eg;\text{vacuum}}$ at $t=0$ to
$-\ket{ge;\text{vacuum}}$ at $t=T$, thereby transferring the excitation
from qubit 1 to qubit 2 while suppressing the link population~\cite{chang2020}. In the multimode and superstrong-coupling regimes, the corresponding state is instead a generalized quasi-dark eigenstate of the full emitter--link system and may contain small photonic components. The spectral gaps identified in Sec.~\ref{sec:non_diff_spectrum} remain open throughout the pulse sequence, permitting quasiadiabatic following of this branch from $\ket{eg;\text{vacuum}}$ to $-\ket{ge;\text{vacuum}}$ across the full passage.}

The STIRAP protocol uses a counterintuitive pulse sequence where the emitter node is activated
\begin{equation}
  \gamma^{\rm STIRAP}_1(t)=\gamma_0\sin^2\!\left(\frac{\pi t}{2T}\right), \quad t\in[0,T],
\end{equation}
while the receiver node is deactivated with some delay $\gamma^{\rm STIRAP}_2(t)=\gamma^{\rm STIRAP}_1(T-t)$. 
For a fixed \rb{dimensionless parameter} $\gamma_0\tau$, the QST error $1-F$ overall tendency is to decrease with increasing $T$, with relevant oscillations induced by the residual bright-state components~\cite{ban2018}.  \rb{Since the first local minimum already corresponds to sufficiently high fidelity, we choose the associated time as the optimal operating time $T_\text{STIRAP}$}.

As shown in Appendix~\ref{app_stirap}, scanning across $\gamma_0\tau$ yields a quadratic error scaling
\begin{equation}
  \label{eq:STIRAP-error}
  \varepsilon\sim 2 \times10^{-5} {(\gamma_0\tau)}^2,
\end{equation}
with optimal duration $T_\text{STIRAP}\sim9\sqrt{\tau/\gamma_0}$. The quadratic suppression of errors reflects the adiabatic protocol’s insensitivity to retardation-induced perturbations: STIRAP’s dark-state trajectory accrues error only at second order in the deviation from perfect adiabaticity, which is itself set by $\gamma_0\tau$.
\rb{As shown in Fig.~\ref{fig:figure_qst}}, STIRAP outperforms SWAP across both the cavity-like and short-link regimes. Even at $\gamma_0\tau=1.44$, \rb{for which the
duration used in Fig.~5 is $T/\tau=6.98$,} the STIRAP infidelity remains below $4\times10^{-4}$, confirming near-ideal performance throughout the entire short-link regime and challenging the traditional view that STIRAP is exclusive to single-mode cavity systems~\cite{serafini2006,chen2007,zhou2009}.

While the STIRAP protocol was derived in the DDE limit, the overall tendency remains the same in full WW simulations with specific detunings $\Delta/\delta\omega_\text{FSR}=50$, $5$, and $1$ (c.f. Fig.~\ref{fig:figure_qst}). The introduction of a lower cutoff in the spectrum only changes the denominator from $5\times 10^{4}$ to a slightly smaller value, confirming the robustness of the STIRAP protocol even in regimes $\Delta\sim \delta\omega_\text{FSR}$ where a more sophisticated treatment of the Lamb shift and of the memory function is required.

\subsection{When synchronization outperforms wavepacket engineering%
\label{sec:czkm}}

The Cirac--Zoller--Kimble--Mabuchi (CZKM) protocol~\cite{cirac1997} is the established approach to quantum state transfer in the long-link (waveguide) regime, where propagation delay dominates and resonant exchange between emitters is unavailable \cite{vogell2017,sinha2020,penas2022,peñas2023,peñas2024,qiu2025,sinha2025,he2025}.
Rather than exploiting synchronization, CZKM engineers the emitted photon wavepacket so that the field exiting qubit~1 is perfectly absorbed by qubit~2 with no reflection back into the link; we include it here as the natural performance baseline to identify the crossover beyond which wavepacket engineering becomes the preferred strategy.
Within our DDE framework, the perfect-absorption condition reads
\begin{equation}
  \sqrt{\gamma_1(t)}c_1(t) + \sqrt{\gamma_2(t+\tau)}c_2(t+\tau)= 0 , 
  \quad \rb{\forall\,t\in[0,T]},
\end{equation}
which is the generalization of the input--output dark-state condition to the short-link regime.
Following Ref.~\cite{penas2022}, we choose a $sech$-like photon wavepacket, yielding the control profile
\begin{equation}
  \gamma^{\rm CZKM}_1(t)=\frac{\gamma_0}{2}\left[1+\tanh\left(\frac{\gamma_0(t-t_c)}{2}\right)\right], \quad t\in[0,T],
\end{equation}
with $\gamma^{\rm CZKM}_2(t)=\gamma^{\rm CZKM}_1(2t_c-t)$, where the $t_c = T/2+\tau/2$ means the center of symmetry.

The DDE framework reveals that CZKM remains viable in the short-link regime ($\gamma_0\tau\ll1$), contrary to the view that it is exclusive to the long-link limit~\cite{he2025}.
\ra{The DDE simulation at $\gamma_0\tau=0.1$, presented in Appendix~\ref{app_czkm} and confirmed by the WW simulation,} shows that retardation effects persist even in this limit, consistent with the single-qubit physics of Sec.~\ref{sec:dtc}, and the protocol achieves high fidelity.
We derive an exact lower bound on the CZKM infidelity
\begin{equation}
  \varepsilon\gtrsim\exp\left[-\gamma_0(T-\tau)\right],
\end{equation}
which depends on the \rb{maximum effective emission rate} $\gamma_0$ and the total duration $T$ as independent parameters. However, specifically for the $sech$-like pulses, we empirically find a larger error
\begin{equation}
  \varepsilon\sim\exp\left[-\gamma_0(T-\tau)/2\right],
\end{equation}
that can be explained by the truncation of the hyperbolic secant tails (see Appendix~\ref{app_czkm}).

Unlike SWAP and STIRAP, where the error is set by $\gamma_0\tau$ alone, CZKM can achieve high fidelity at any $\gamma_0\tau$ by extending $T$, at the cost of a longer protocol.
Comparing the protocols under identical resource constraints ($\gamma_0$, $T$) (Fig.~\ref{fig:figure_qst}), we find a crossover at $\gamma_0\tau\approx1.44$, $T\approx6.98\tau$: \rb{in the lower-$T/\tau$ range, $T/\tau<6.98$, CZKM yields a lower
infidelity than STIRAP, whereas for $T/\tau>6.98$, STIRAP performs better.}

\subsection{Sensitivity to photon loss in the link%
\label{sec:losses}}
\begin{figure}[t]
  \centering \includegraphics[width=\columnwidth]{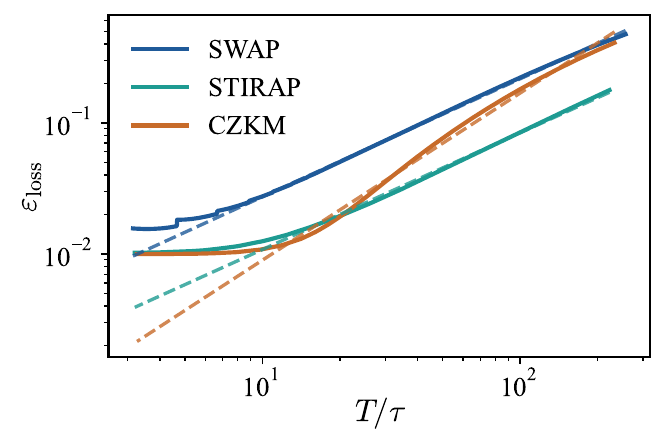}
  \caption{
    Photon loss in the link. \rb{Solid lines show the estimated loss 
    $\varepsilon_{\rm loss}$ as a function of $T/\tau$ at fixed $\kappa\tau=0.01$ for the three 
    protocols (SWAP, STIRAP, and CZKM)}, while dashed lines of the same 
    color indicate the corresponding power-law fits. 
    The fitted scalings are 
    $\varepsilon_{\rm loss} = 0.0034\,(T/\tau)^{0.9}$ (SWAP), 
    $0.0014\,(T/\tau)^{0.9}$ (STIRAP), and 
    $0.0005\,(T/\tau)^{1.3}$ (CZKM).
    \label{fig:losses}}
\end{figure}
If our link loses photons with a global loss rate $\kappa$, our quantum state will deviate from the Wigner-Weisskopf ansatz, introducing some probability of the experiment ending with zero excitations. We can quantify the probability of this happening as a multiplicative contribution to the infidelity, giving an error
\begin{equation}
  \label{eq:ploss}
  \varepsilon_{\rm loss} = 1 - \exp\!\left(-\kappa \int_0^T n(t)\,dt\right).
\end{equation}
In this formula, $n(t)$ is the number of photons in the link. This quantity may be computed from the numerical simulations of our protocols.

Our main conclusion, using a pessimistic estimate $\kappa\tau=0.01$, is the superiority of the STIRAP protocol from the point of view of losses (c.f. Fig.~\ref{fig:losses}). This can be explained by the fact that STIRAP protocols keep the link weakly populated throughout the transfer~\cite{vogell2017}, while the SWAP and CZKM protocols rely on the creation of one flying photon.

Note also, that the improved losses in STIRAP can enlarge the range over which this protocol becomes more favorable than CZKM, reaching slightly above $\gamma_0\tau=1.44$. However, this is more of a setup-specific consideration.

\section{Conclusion}
The Delay Differential Equation (DDE) framework introduced here allows us to systematically benchmark three quantum state transfer protocols across the full $\gamma_0\tau$ parameter space, with predictions that survive for finite values of $\Delta/\delta\omega_\text{FSR}$.
STIRAP, exploiting quasi-dark states, achieves a quadratic error floor $\mathcal{O}(2\times10^{-5}{(\gamma_0\tau)}^2)$ and outperforms SWAP throughout the short-link regime.
Photon-shaped CZKM becomes optimal beyond the crossover $\gamma_0\tau\approx1.44$, as quantified by the analytical infidelity bound that applies to all CZKM protocols.

These protocols are enabled by a photon-echo-driven synchronization in short quantum links: emitters spontaneously lock into piecewise non-differentiable Rabi oscillations at frequency $\Omega_R\approx\sqrt{\gamma/2\tau}$, driven by coherent echoes returning after each round trip $2\tau$, constituting an autonomous discrete time-crystal-like phenomenon in the single-excitation sector.
The Rabi exchange drives SWAP, while the persistent quasi-dark states---confirmed by spectral analysis across the full cavity-to-waveguide crossover---sustain STIRAP.
The DDE framework provides a unified, exact description of this regime, yielding closed-form analytical solutions for both the dynamics and the spectral structure.

The self-synchronization kinks are experimentally accessible via time-resolved field detection: the derivative discontinuities at $t=n\cdot2\tau$ in the emitter excited-state population manifest as abrupt changes in the emitted photon field via the input--output relation $b^{\rm out}(t)\propto\sqrt{\gamma}\,c(t)$. At $\gamma\tau\approx0.2$ the kink amplitude is bounded by $2\gamma$; platforms with larger $\gamma\tau$, such as the 5~m link of Magnard \textit{et al.}~\cite{magnard2020} ($\gamma\tau\approx 1$) and slow-light waveguides that extend the effective travel time~\cite{ferreira2024}, provide more favorable signal conditions for a first observation.

Spectroscopic confirmation follows equally directly from sweeping the qubit frequency across a resonance of the link and measuring the avoided crossings and quasi-dark-state suppressions of Fig.~\ref{fig:power_spectrum}.
These results map directly onto current circuit-QED hardware: the 0.73~m link of Chang \textit{et al.}~\cite{chang2020} ($\gamma\tau\approx0.2$) and the 5~m link of Magnard \textit{et al.}~\cite{magnard2020} ($\gamma\tau\approx 1$) lie squarely in the STIRAP-optimal regime, while the 64~m cryogenic cable of Qiu \textit{et al.}~\cite{qiu2025} ($\gamma\tau\approx2$) enters the CZKM-optimal, retardation-dominated regime.

The short-link regime is not a trivial transition zone between cavity and waveguide QED but a physically rich parameter space where photon-echo synchronization becomes an engineering resource---one for which DDE modeling provides the exact analytical foundations needed to design and optimize quantum link experiments. We anticipate that the DDE framework, which complements recent investigations based on tensor network formalisms~\cite{vodenkova2024, arranzregidor2025, capurso2025} and operator DDE~\cite{barahona-pascual2025},
will be a powerful tool for exploring and exploiting the physics of quantum links and larger quantum optical networks, enabling discovery of novel phenomena and development of new protocols that leverage the unique features of this regime.

\begin{acknowledgments}
  This work has been supported by CSIC' JAE Chair Program 2024 and PRO-ERC AGAIN 2025 funding. J.J.G.R. acknowledges support by grant NSF PHY-2309135 to the Kavli Institute for Theoretical Physics (KITP).
  H.J. gratefully acknowledges financial support from the program of China Scholarship Council (CSC202308620117). C.B. gratefully acknowledges financial support from the Comunidad de Madrid via the PIPF-2024 grant (COM-34268).
\end{acknowledgments}
\section*{DATA AVAILABILITY}
\rd{The numerical codes used in this work are publicly available at Zenodo~\cite{zenodo_code}.}

\appendix
\section{two-qubit DDE derivation}
We apply the Hamiltonian in Eq.~\eqref{eq:Hamiltonian} to the ansatz in Eq.~\eqref{eq:ww_ansatz}, which yields
\begin{align}
    \dot{c}_l(t) =& -i\Delta c_l(t) - i \sum_k g_{kl}(t) \alpha_k(t) \\
    \dot{\alpha}_k(t) =& -i\omega_k \alpha_k(t) -i\sum_j g_{k,j}^\ast(t) c_{j}(t)
\end{align}
where $g_{kl}(t)=g_l(t)\cos(kx_l)$. Here, $g_l(t)$ denotes the time-dependent coupling strength between qubit-$l$ and the waveguide, whereas $\cos(kx_l)$ represents the position-dependent mode-function factor arising from the propagation phase at the qubit position $x_l$.
And then the last equation can be formally integrated 
\begin{equation}
    \alpha_k(t) = \alpha_k(0)e^{-i\omega_kt}-i \int_{0}^{t} \sum_{j=1}^{2} g_{k,j}^\ast(u) e^{-i\omega_k (t-u) }c_{j}(u) du
\end{equation}
which is then inserted into the equation for $c_{l}(t)$, resulting in 
\begin{align}
    \dot{c}_l(t) =&  -i\Delta c_{l}(t) - i\sum_k g_{kl}(t)\alpha_k (0) e^{-i\omega_kt } \\
    - &\int_0^t \sum_{j=1}^{2} K_{lj}(t,u) c_j(u) du
\end{align}
This last term is the effective emitter-emitter interaction, weighted by a kernel function 
\begin{equation}
    K_{lj}(t,u) = \sum_k g_{kl}(t)g_{kj}(u)^\ast e^{-i\omega_k(t-u)} .
\end{equation}
The full kernel $K_{lj}(t,u)$ retains a joint dependence on its two time arguments through the coupling envelopes $g_l(t)g_j^\ast(u)\propto\sqrt{\gamma_l(t)\gamma_j(u)}$. Only the bath correlation arising from the mode sum depends on the relative time $t-u$. In the broadband limit this correlation becomes a sum of delta functions in $t-u$: a local contribution proportional to $\delta(t-u)$ together with the propagation echoes of the closed line. The slowly varying coupling envelopes remain outside these delta functions and are therefore evaluated at the present time $t$ and at the corresponding retarded time $u$, as shown explicitly in Eqs.~(A7) and~(A8).

For a linear dispersion relation, we have $\omega_k = \omega_m = m \omega_{\text{fsr}}$, with $\omega_{\text{fsr}} = \frac{\pi c}{L} = \frac{\pi}{\tau}$. Assuming that the coupling strengths vary slowly in time and that the waveguide bandwidth is sufficiently broad, we consider two qubits located at the two ends of the waveguide, with $x_1=0$ and $x_2=L$. For convenience, we further assume that $g_l(t)$ is real and positive, i.e., $g_l(t)\in\mathbb{R}_{\geq0}$.
The corresponding kernels are
\begin{align}
   K_{ll}(t,u)&=\sum_{m=-\infty}^{\infty} \sqrt{\gamma_l(t)\gamma_l(u)}\, \delta(t-u-\tau_1^m), \\
   K_{lj}(t,u)&=\sum_{m=-\infty}^{\infty} \sqrt{\gamma_l(t)\gamma_j(u)}\, \delta(t-u-\tau_2^m), \qquad j\ne l,
\end{align}
where we use the standard input--output normalization $|g_l(t)|^2=\gamma_l(t)$, with the density-of-states factors absorbed into the effective emission rate $\gamma_l(t)$~\cite{barahona-pascual2025}, and where $\tau_1^m=2m\tau$ and $\tau_2^m=(2m+1)\tau$. If the field's initial state is the vacuum, $\alpha_k(0)=0$ and the equations simplify to 
\begin{align}
    \dot{c}_l(t) = &  -i\Delta c_l(t) - \frac{\gamma_l(t)}{2} c_l(t)  \\
    & - \sqrt{\gamma_l(t) } \sum_{m=1}^{\infty}\sqrt{\gamma_{l}(t-\tau_1^m) } c_{l}(t-\tau_1^m) \Theta(t-\tau_1^m)  \\  
    & \!\!\!\!\!\!\!\!\!\!\!\!\!\!\!\!\!\!-\sqrt{\gamma_l(t)}\sum_{m=0}^{\infty}\sqrt{\gamma_{3-l}(t-\tau_2^m) } c_{3-l}(t- \tau_2^m) \Theta(t - \tau_2^m)
\end{align}
Upon transforming to a frame rotating at the qubit frequency and collecting the retarded contributions into the output fields $b_l^{\mathrm{out}}$, while retaining the time-dependent emission rates $\gamma_l(t)$, we recover Eq.~\eqref{eq:dde_two_emitter} in the main text:
 \begin{align}
  \frac{d c_{l}(t)}{dt}  =& -\frac{\gamma_{l}(t)}{2}\,c_{l}(t)  \,   \\
  &-\sqrt{\gamma_l(t)} \big[e^{i2\phi}b_{l}^\text{out}(t-2\tau)  +
  e^{i\phi}  b_{3-l}^{\text{out}}(t-\tau) \big], \notag
\end{align}
where $\phi=\Delta\tau$, and
\begin{equation}
  b_{l}^\text{out}(t) =\sum_{n=0}^{\infty}e^{i2n\phi}\left.\sqrt{\gamma_l(t_n)} \,  c_{l}(t_n)\Theta(t_n)\right|_{t_n=t-2n\tau}.
\end{equation}
\section{Solution of the Single-Qubit DDE}
\label{app_analytical_dynamics}
We show how the dynamics of a single qubit in a quantum link can be described as a polynomial series multiplied by an exponential factor.
We solve  for a single qubit, using $\gamma_{l}(t) = \gamma \delta_{l,1}$ and $c_{l}(t) =c(t)\delta_{l,1}$. Performing the change of variables $c(t) = e^{\frac{-\gamma t}{2}} f(t) $ yields
\begin{equation}
  \frac{d}{dt}f(t)  = -\gamma \sum_{n=1}^{\infty}e^{n(i\phi+\frac{\gamma \tau }{2})}f(t-n\tau)\Theta(t-n\tau). \label{eq:scalar_simple_eq}
\end{equation}
We use the Laplace-transformed version of the equation, with $\mathcal{F}(s) = \mathcal{L}\left[ f\right](s)$ the Laplace transform,  our DDE becomes
\begin{equation}
  s\mathcal{F}(s)- f(0) = -\gamma \left[-1+\sum_{n=0}^{\infty}e^{n\left(i\Delta + \frac{\gamma}{2}\right)\tau}e^{-ns\tau} \right]\mathcal{F}(s),
\end{equation}
where we have used $\phi = \Delta\tau$. Let us call $\alpha = i\Delta+ \frac{\gamma}{2}$. Using the geometric series, we get
\begin{equation}
  \mathcal{F}(s) = \frac{1}{s} \left[ \frac{1- e^{(\alpha-s)\tau}}{1- e^{(\alpha-s)\tau} \left(1-\frac{\gamma}{s} \right)} \right].
\end{equation}
Once we expand the denominator using the geometric series again, we use the binomial formula to get
\begin{equation}
  \mathcal{F}(s)  = \frac{1- e^{(\alpha-s)\tau}}{s} \sum_{n=0}^{\infty} {
  \left(  e^{(\alpha-s)\tau} \right)}^{n}
  \sum_{m=0}^{n}\binom{n}{m} {\left(\frac{-\gamma}{s} \right)}^{m}.
\end{equation}
After some manipulation, this can be converted into
\begin{equation}
  \mathcal{F}(s) = \frac{1}{s} + \sum_{n=1}^{\infty} {\left( e^{ \alpha\tau} \right)}^{n} \sum_{m=1}^{n} \binom{n-1}{m-1}{(-\gamma)}^{m} \frac{e^{-ns\tau}}{s^{m+1}} ,
\end{equation}
and, finally, making the inverse Laplace transform results in
\begin{align}
  f(t) =&  \left\{ 1 + \sum_{n=1}^{\infty}e^{n\alpha \tau} \sum_{m=1}^{n} \mathcal{P}_{n,m}(t) \Theta(t-n\tau)\right\} \\
  \mathcal{P}_{n,m}(t) &=\binom{n-1}{m-1} \frac{[-\gamma (t-n\tau)]^{m}}{m!}
\end{align}
Inverting the change of variables gives the solution we were seeking.

\section{Upper Bound on the derivative's Discontinuities}
\label{app_discontinuity_bound}
The discontinuities in the derivative occur when $t=N\tau$. We can use the DDE's structure to analyze these discontinuities, by making small perturbations. In particular, we can find
\begin{equation}
  \lim_{\delta t\to 0}\left[ \partial_{t}c(N\tau) - \partial_{t}c(N\tau-\delta t ) \right] = -\gamma e^{iN{\phi}}c(0),
\end{equation}
Since $|c(0)|\leq 1$, we can establish that the discontinuities in the derivative of $c(t)$ are upper-bounded by $\gamma$.

The generalization to the excited-state population $|c(t)|^{2}$ follows the same approach. Using $|c(t)|^{2} = c^{\ast}(t)c(t)$ we get
\begin{align}
  &  \left | \lim_{\delta t \to 0} \left[ \partial_{t}|c(N\tau)|^{2}- \partial_{t}|c(N\tau-\delta t)|^{2}\right] \right | = \\
  & =\left|-2\gamma\, \text{Re} \left\{ e^{iN\phi}c^{\ast}(N\tau)c(0)\right\} \right|  \leq2\gamma,
\end{align}
where we use the fact that $c(t) \leq 1$.

\section{Computation of eigenfrequencies}
\label{app_eigenstates}
We can use input-output theory to compute the field operator $a_{\text{out}}(t)$ in the cavity. To the standard relation $a_{\text{out}}(t) = a_{in}(t) + \sqrt{\gamma} c(t)$ we add the condition arising from the closed ends from the cavity, 
\begin{equation}
    a_{\text{in}}(t) = e^{i\phi}a_{\text{out}}(t-\tau),
\end{equation}
which means we can write 
\begin{equation}
    a_{\text{out}}(t) =  e^{i\phi}a_{\text{out}}(t-\tau) + \sqrt{\gamma}c(t). 
\end{equation}
By performing a Laplace transform , this is just 
\begin{equation}
     \mathcal{A}_{\text{out}}(s) = \frac{\sqrt{\gamma}}{1-e^{i\phi-s\tau}}\mathcal{C}(s)
\end{equation}
The Laplace transform $\mathcal{C}(s)$ can be obtained from the DDE directly, and, after some manipulation, we get 
\begin{equation}
      \mathcal{A}_{\text{out}}(s)   = \frac{\sqrt{\gamma} }{(s+\frac{\gamma}{2})(1-e^{i\phi -s\tau}) + \gamma e^{i\phi -s\tau}} . 
\end{equation}
We make the change in variables $s = -i\omega$ and look at the poles of $\mathcal{A}_{\text{out}}(\omega)$, which gives the equation 
\begin{equation}
   -i\omega+\frac{\gamma}{2} + \gamma \frac{e^{i(\phi +\omega\tau)}}{1-e^{i(\phi +\omega\tau)}} = 0 .
\end{equation}
After some manipulation, it results in the trascendental equation :
\begin{equation}
    \omega =\frac{\gamma}{2}\cot\left(\frac{\omega \tau + \phi }{2}  \right),
\end{equation}
which only has solutions with in the real plane, meaning that those are eigenfrequencies with no decay. Finally, one can revert the rotating frame by displacing the energies $\omega \to \omega-\Delta $, resulting in the equation presented in the main text.

\section{Quantum state transfer via SWAP%
\label{app_swap}}
Under constant qubit-waveguide couplings, the ideal single-mode cavity QED description predicts optimal quantum state transfer at $\Omega T=\pi$.
In the short-link regime, multimode effects and propagation delay shift the optimal transfer point away from this condition.

\begin{figure}[t!]
  \centering
  \includegraphics[width=\columnwidth]{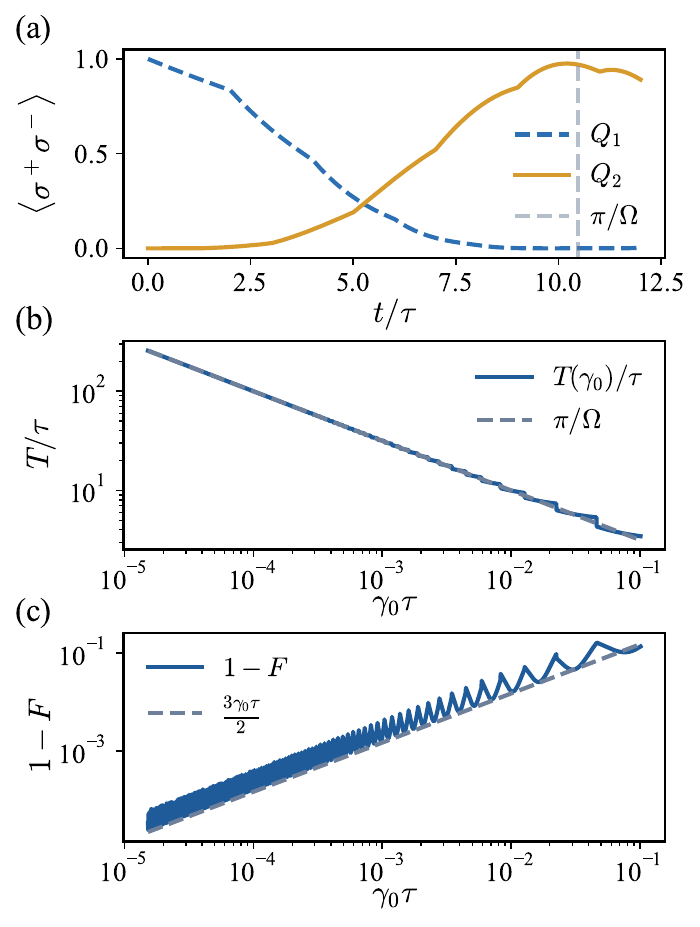}
  \caption{
    Dynamics and transfer efficiency of the SWAP protocol.
    (a) Typical system dynamics at a fixed \rb{dimensionless parameter} $\gamma_0\tau=0.09$.
    The vertical dashed line indicates the ideal Rabi period $T=\pi/\Omega$.
    (b) Optimal transfer time $T(\gamma_0)$ as a function of the \rb{maximum effective emission rate}.
    The orange line denotes the baseline given by the Rabi resonance condition.
    (c) Minimum transfer error $1-F$ (solid line) versus the \rb{dimensionless parameter $\gamma_0\tau$}.
    The dash line shows a linear fit $1-F \geq \frac{3}{2}\gamma_0\tau$,
    illustrating that the optimal SWAP infidelity increases approximately linearly with \rb{$\gamma_0\tau$}.
  }%
  \label{fig:swap_app}
\end{figure}

Results from numerical optimization are shown in Fig.~\ref{fig:swap_app}.
At fixed $\gamma_0\tau$, the fidelity peak deviates from the ideal Rabi time $T=\pi/\Omega$, as seen in  Fig.~\ref{fig:swap_app}a.
The optimized transfer time (Fig.~\ref{fig:swap_app}b) fluctuates around $T=\pi/\Omega$, accompanied by oscillations in the infidelity due to multimode interference.

The minimal infidelity follows a linear scaling with $\gamma_0\tau$(Fig.~\ref{fig:swap_app}c):
\begin{equation}
  1-F \gtrsim \tfrac{3}{2}\gamma_0\tau.
\end{equation}
This linear scaling characterizes the performance limit of the SWAP protocol in the short-link regime.

\section{Quantum state transfer via STIRAP%
\label{app_stirap}}

\begin{figure}[t!]
  \centering
  \includegraphics[width=\columnwidth]{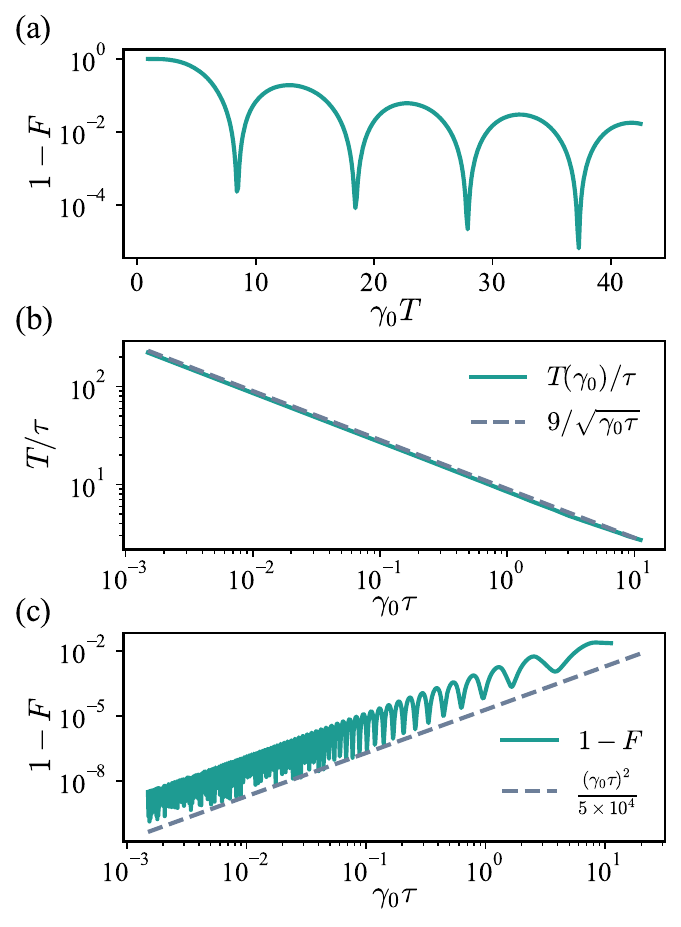}
  \caption{
    Dynamics and transfer efficiency of the STIRAP protocol.
    (a) Infidelity $1-F$ as a function of the protocol duration $T$ at a fixed \rb{dimensionless parameter} $\gamma_0\tau$.
    (b) Optimized transfer time $T(\gamma_0)$ as a function of \rb{$\gamma_0\tau$}.
    (c) Minimum transfer infidelity $1-F$ (solid line) versus $\gamma_0\tau$.
    The dash line represents a power-law fit given by $1-F \simeq \frac{(\gamma_0\tau)^2}{50000}$,
    illustrating that the STIRAP infidelity scales approximately quadratically with \rb{$\gamma_0\tau$}.
  }%
  \label{fig:stirap_app}
\end{figure}

We numerically optimize the optimal quantum state transfer (QST) position for the STIRAP-like protocol, whose performance is summarized in Fig.~\ref{fig:stirap_app}.

At a fixed value of $\gamma_0\tau$, the QST error decreases with the protocol duration $T$ while exhibiting oscillatory behavior, as shown in Fig.~\ref{fig:stirap_app}a. Due to the adiabatic nature of the protocol, the overall fidelity decreases continuously over time, and the oscillations arise from the accumulated phase of the residual bright state.

For each value of the \rb{dimensionless parameter $\gamma_0\tau$}, we optimize $T$ to achieve the highest fidelity at the first infidelity $1-F$ valley, which we define as the optimal QST position. The corresponding optimal transfer time and minimal infidelity are presented in Fig.~\ref{fig:stirap_app}b and Fig.~\ref{fig:stirap_app}c, respectively.

The infidelity follows a quadratic scaling relation with the \rb {dimensionless parameter} $\gamma_0\tau$, and the corresponding fitting asymptote is given by Eq.~\eqref{eq:STIRAP-error}.
This fitting asymptote indicates that even when $\gamma_0\tau \sim 1$ (short-link regime), the QST error remains as low as approximately $10^{-5}$.

\section{Quantum state transfer via CZKM%
\label{app_czkm}}

\begin{figure}[t!]
  \centering
  \includegraphics[width=\columnwidth]{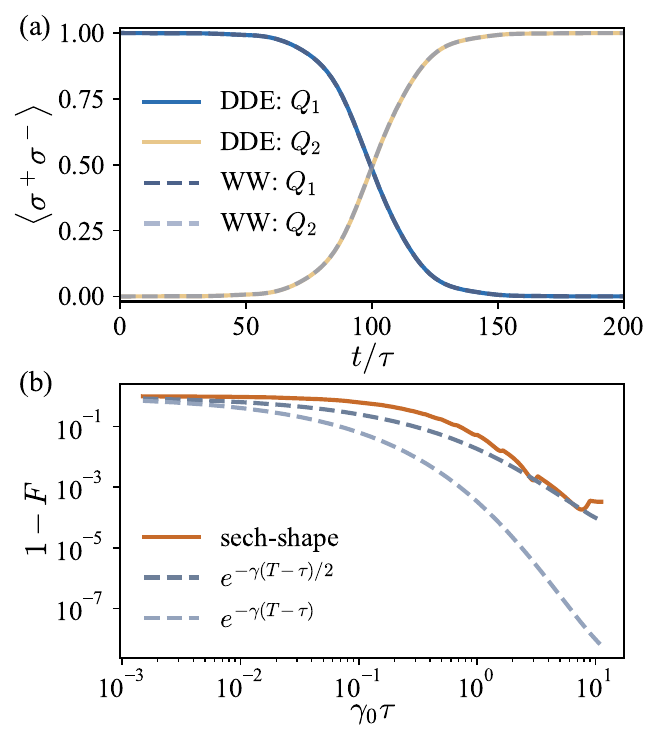}
  \caption{QST error and dynamics for the CZKM protocol. 
  (a) Population dynamics for the same $sech$-like photon-shaping scheme at $\gamma_0\tau=0.1$ and $T=200\tau$.\ra{$N_{\mathrm{modes}}=201$ for the WW benchmark points.}
  (b) QST error. The solid curve shows exact DDE simulations using a $\mathrm{sech}$ photon-shaping scheme under the same resource constraint as STIRAP ($T=9/\sqrt{\gamma_0 \tau}$). 
    The two dashed lines denotes the approximate $\mathrm{sech}^2$ prediction and the ideal dark-state bound. }%
  \label{fig:CZKM_app}
\end{figure}

In the photon-shaping protocol, a target wave packet $\psi(t)$ is generated by controlling the emitter coupling $\gamma_1(t)$. Under the ideal dark-state condition, the emitter amplitude satisfies
\begin{align}
  \dot c_1(t)=-\frac{\gamma_1(t)}{2}c_1(t),
\end{align}
with solution $|c_1(t)|^2=\exp[-\int_{t_0}^{t}\gamma_1(t')dt']$. The emitted photon is $\psi(t)=-i\sqrt{\gamma_1(t)}c_1(t)$, which leads to
\begin{align}
  \gamma_1(t)=\frac{|\psi(t)|^2}{1-\int_{t_0}^{t}|\psi(t')|^2dt'}.
  \label{eq:shape_general_app}
\end{align}

For an infinitely long normalized $\mathrm{sech}$ wave packet,
\begin{align}
  |\psi(t)|^2=\frac{\gamma_0}{4}\,\mathrm{sech}^2\!\left(\frac{\gamma_0 t}{2}\right),
\end{align}
one obtains
\begin{align}
  \gamma_1(t)=\frac{\gamma_0}{2}\left[1+\tanh\!\left(\frac{\gamma_0 t}{2}\right)\right],
  \label{eq:gamma_sech_app}
\end{align}
with $\bar\gamma_2(t)=\gamma_2(t+\tau)=\gamma_1(-t)$.

For a finite protocol duration, however, the exact dark-state condition cannot be maintained throughout the evolution. We therefore derive the QST error directly from the full DDE dynamics.

Introducing the shifted variable $\bar c_2(t)=c_2(t+\tau)$, we define the dark and bright amplitudes
\begin{align}
d(t)&=\frac{\sqrt{\bar\gamma_2(t)}\,c_1(t)-\sqrt{\gamma_1(t)}\,\bar c_2(t)}{\sqrt{\gamma_0}},
\\
b(t)&=\frac{\sqrt{\gamma_1(t)}\,c_1(t)+\sqrt{\bar\gamma_2(t)}\,\bar c_2(t)}{\sqrt{\gamma_0}}.
\end{align}
For the $\mathrm{sech}$ protocol, one finds $\dot d(t)=0$, while
\begin{align}
\dot b(t)=
-\frac{\gamma_0}{2}b(t)
-\gamma_0\sum_{n=1}^N b(t-2n\tau)\,\Theta(t-2n\tau).
\label{eq:b_dde_final}
\end{align}

We define the QST duration as $[-T_{\mathrm{eff}}/2,\,T_{\mathrm{eff}}/2]$, where $T_{\mathrm{eff}}=T-\tau$. With initial conditions $c_1(-T_{\mathrm{eff}}/2)=1$ and $\bar c_2(-T_{\mathrm{eff}}/2)=0$, we introduce $u=\tanh(\gamma_0 T_{\mathrm{eff}}/4)$, yielding $d_0=\sqrt{(1+u)/2}$ and $b_0=\sqrt{(1-u)/2}$.

The final transfer amplitude can be written as
\begin{align}
\bar c_2(T_{\mathrm{eff}})=
-\frac{1+u}{2}
+\frac{1-u}{2}\,\beta(T_{\mathrm{eff}}),
\end{align}
where $\beta(t)=b(t)/b_0$ satisfies
\begin{align}
\dot\beta(t)=
-\frac{\gamma_0}{2}\beta(t)
-\gamma_0\sum_{n=1}^N \beta(t-2n\tau)\,\Theta(t-2n\tau),
\end{align}
with $\beta(-T_{\mathrm{eff}}/2)=1$.

The exact QST error is therefore
\begin{align}
\varepsilon(T_{\mathrm{eff}},\tau)
=
1-
\left|
\frac{1+u}{2}
-
\frac{1-u}{2}\,\beta(T_{\mathrm{eff}}
)
\right|^2.
\label{eq:error_final_app}
\end{align}

For $\gamma_0 T_{\mathrm{eff}}\gg1$, using $u=1-2e^{-\gamma_0 T_{\mathrm{eff}}/2}+O(e^{-\gamma_0 T_{\mathrm{eff}}})$, Eq.~\eqref{eq:error_final_app} reduces to
\begin{align}
\varepsilon(T_{\mathrm{eff}},\tau)
&=
2e^{-\gamma_0 T_{\mathrm{eff}}/2}\bigl[1+\mathrm{Re}\,\beta(T_{\mathrm{eff}})\bigr]
+O(e^{-\gamma_0 T_{\mathrm{eff}}})\notag\\
&\sim e^{-\gamma_0 T_{\mathrm{eff}}/2}.
\label{eq:error_asymp_app}
\end{align}
Thus, the leading error still scales as $e^{-\gamma_0 T_{\mathrm{eff}}/2}$, while the prefactor is renormalized by the bright-state dynamics encoded in $\beta(T_{\mathrm{eff}})$.

Furthermore, in the ideal dark-state limit, probability conservation implies $\frac{d}{dt}[|c_1(t)|^2+|\bar c_2(t)|^2]=0$, so that $|\bar c_2(T_{\mathrm{eff}})|^2 = 1-|c_1(T_{\mathrm{eff}})|^2$. Using the solution of $c_1(t)$, one obtains
\begin{align}
  F = 1-\exp\!\left[-\int_{-T_{\mathrm{eff}}/2}^{T_{\mathrm{eff}}/2}\gamma_1(t')dt'\right].
\end{align}
For any protocol with $\gamma_1(t)\le\gamma_0$,
\begin{align}
  1-F \ge \exp\!\left[-\gamma_0 T_{\mathrm{eff}}\right].
  \label{eq:dark_bound_app}
\end{align}

These results are shown in Fig.~\ref{fig:CZKM_app}b. The exact DDE results closely follow the $\mathrm{sech}$-like pulse prediction, 
while both remain above the ideal dark-state bound, indicating that finite-duration effects primarily renormalize the prefactor 
without altering the exponential scaling.

\bibliography{usc_biblio}

\end{document}